\documentclass[seceq]{ptptex}    
\usepackage{amsmath,amsfonts,latexsym,amssymb} 
\usepackage{verbatim,amsthm,graphics} 
\usepackage{graphicx}

\DeclareFontFamily{U}{rsfs}{\skewchar\font"7F}  
\DeclareFontShape{U}{rsfs}{m}{n}{<-6> rsfs5 <6-8> rsfs7 <8-> rsfs10}{}
\DeclareMathAlphabet{\mathscr}{U}{rsfs}{m}{n}

\def\SHB{{\mathbb S}}

\def\scalar{{\mathbb S}} 
\def\vector{{\mathbb V}} 
\def\tensor{{\mathbb T}}

\def\ben{\begin{equation}} 
\def\een{\end{equation}}
\def\bena{\begin{eqnarray}}
\def\eena{\end{eqnarray}}

\def\f(#1/#2){\frac{#1}{#2}} 
\def\Frac(#1/#2){\left(\frac{#1}{#2}\right)}

\newcommand{\non}{\nonumber}

\begin{document}

\title{Stability of Higher-Dimensional Schwarzschild Black Holes}  
\author{ Akihiro {\sc Ishibashi}$^{1,}$%
\footnote{E-mail: A.Ishibashi@damtp.cam.ac.uk} 
and Hideo {\sc Kodama}%
$^{2,}$%
\footnote{E-mail: kodama@yukawa.kyoto-u.ac.jp} 
}

\inst{$^1$D.A.M.T.P., Centre for Mathematical Sciences\\
  University of Cambridge, \\
  Wilberforce Road, Cambridge CB3 0WA, United Kingdom \\ 
$^2$Yukawa Institute for Theoretical Physics\\
Kyoto University, Kyoto 606-8502, Japan 
\\
(Received June 3, 2003)  
\\ 
} 

\abst{ 
We investigate the classical stability of higher-dimensional 
Schwarzschild black holes with respect to linear perturbations 
in the framework of a gauge-invariant formalism for gravitational 
perturbations of maximally symmetric black holes.  
This formalism was recently developed by the present authors. 
The perturbations are classified into three types, those of 
tensor, vector, and scalar modes, according to their tensorial 
behaviour on the spherical section of the background metric. 
The vector- and scalar-type modes correspond, respectively, 
to the axial and polar modes in the four-dimensional case. 
We show that for each mode of the perturbations, the spatial
derivative part of the master equation is a positive, self-adjoint 
operator in the $L^2$-Hilbert space, and 
hence that the master equation for each type of perturbation 
has no normalisable negative modes that would correspond to 
unstable solutions.  
\par
In the same Schwarzschild background, we also analyse 
static perturbations of the scalar mode and show that 
there exists no static perturbation that is regular everywhere 
outside the event horizon and is well behaved at the spatial infinity.  
This confirms the uniqueness of the higher-dimensional spherically 
symmetric, static vacuum black hole, within the perturbation framework. 
\par
Our strategy for treating the stability problem is also applicable 
to other higher-dimensional maximally symmetric black holes 
with a non-vanishing cosmological constant. 
We show that all possible types of maximally symmetric black holes 
(including the higher-dimensional Schwarzschild-de Sitter 
and Schwarzschild-anti-de Sitter black holes) are stable 
with respect to tensor and vector perturbations. 
}

\maketitle

\section{Introduction} 
\label{sect:1}  

Higher-dimensional theories of gravity have recently attracted 
much attention, and black hole solutions have played an 
important role in revealing various aspects of gravity 
peculiar to higher-dimensional 
systems\cite{MyersPerry,Witten1998AdSCFTthermal, 
DimopoulosLandsberg2001,GiddingsThomas2001}.  
A noteworthy fact is that, 
compared to the severely constrained situation 
in four spacetime dimensions\cite{Israel1967,Carter1971,Robinson1975}, 
higher-dimensional black holes can have many very different 
varieties\cite{GibbonsIdaShiromizu2002a,GibbonsIdaShiromizu2002b,%
EmparanReall2002,GibbonsHartnoll2002,GibbonsHartnollPope2003}.   
For this reason, it is important to study the dynamics of 
fields\cite{HorowitzHubeny,CardosoDiasLemos03,Konoplya03,
Molina03,IdaUchidaMorisawa2002} 
and metric perturbations\cite{GibbonsHartnoll2002, 
GibbonsHartnollPope2003,Hartnoll2003} 
in such black hole backgrounds and to find stable solutions. 
Amongst the various types of black hole solutions, 
a natural higher-dimensional generalisation of the Schwarzschild 
metric---also known as the Schwarzschild-Tangherlini 
metric\cite{Tangherlini1963}---has been assumed to be   
(or believed to be) stable, like its four-dimensional counterpart, 
and for this reason it has been conjectured that this metric 
describes a possible final equilibrium 
state of classical evolution of an isolated system, or 
a certain class of unstable black holes in higher dimensions.   
However, the question of the stability of a Schwarzschild black hole 
itself is not a trivial matter, and, to our knowledge, 
it has not yet been fully analysed in higher dimensions.

In the four-dimensional case, the perturbation analysis of 
the Schwarzschild metric was first carried out by Regge and 
Wheeler\cite{ReggeWheeler1957}, and its stability was 
essentially confirmed by Vishveshwara\cite{Vishveshwara1970}, 
followed by further detailed studies\cite{Price,Wald1978}. 
The basic observation of those works is that 
unstable perturbations, whose growth in time is unbounded, 
are initially divergent at the event horizon (provided that 
they are vanishing at large distances) and therefore 
are physically unacceptable.

A key step in the analysis is the reduction of the perturbation    
equations to a simple, tractable form. For this purpose, 
in the case of a four-dimensional spacetime, the spherical symmetry 
of the background Schwarzschild metric is utilised to classify 
the metric perturbations into two types, axial (odd) modes  
and polar (even) modes, 
according to their behaviour under the parity transformation 
on the two-sphere. To linear order, the perturbation equations 
for these two types of perturbations are independent, 
and therefore they reduce to a set of two Schr\"odinger-type 
2nd-order ordinary differential equations (ODEs), 
the Regge-Wheeler equation for axial perturbations  
and the Zerilli equation\cite{Zerilli1970,Zerilli1970b} 
for polar perturbations.  

In the higher-dimensional case, 
the role of the two-sphere is played by an $n$-sphere: 
According to their tensorial behaviour on the $n$-sphere, 
there are three types of perturbations, 
{\em tensor}, {\em vector}, and {\em scalar} perturbations. 
The first of these is a novel type of perturbation that exists 
only in the higher-dimensional case, 
while the last two types correspond, 
respectively, to the axial and polar modes 
in the four-dimensional case.

In order to verify the stability of higher-dimensional Schwarzschild 
black holes, one must investigate their stability with respect to 
all the types of perturbations mentioned above. 
However, as far as the present authors know, 
all analysis made to this time has been carried out for only 
tensor perturbations\cite{GibbonsHartnoll2002}, and 
the situation for vector and scalar perturbations has not yet 
been investigated: The stability of Schwarzschild black holes 
in higher dimensions is not yet established.       
The treatment of scalar perturbations\footnote{ 
 The term ``scalar perturbation'' is often used in reference 
 to a perturbation of a free scalar field in works involving 
 quasi-normal mode analysis of black holes. 
 This is conceptually quite different from the scalar perturbations 
 considered in the present paper, 
 which are modes of metric perturbations 
 of a vacuum spacetime. Furthermore, the effective potentials for 
 these two cases are different. Hence, although the quasi-normal 
 mode analysis of a test scalar 
 field\cite{HorowitzHubeny,CardosoDiasLemos03,Konoplya03,Molina03} 
 may shed some light on the stability problem of black holes 
 in higher dimensions, 
 an exact analysis of the problem based on the formulation 
 given in the present paper is necessary.    
} 
is the most difficult part of this analysis, because 
the linearised Einstein equations for a scalar perturbation 
are much more complicated than those for a tensor perturbation 
and do not reduce to  
the form of a single 2nd-order Schr\"odinger-type ODE 
in the higher-dimensional case. 

Recently, this reduction was carried out by 
the present authors\cite{KI2003} in the background of 
maximally symmetric black holes with a non-vanishing 
cosmological constant in an arbitrary number of spacetime dimensions.    
From the result of that work, along with the previously 
obtained ODEs for tensor 
and vector perturbations\cite{KIS2000}, we have the full set 
of equations for perturbations of higher-dimensional maximally symmetric 
black holes in the form of a single self-adjoint 2nd-order 
ODE---which we refer to as the {\em master equation}---for 
each tensorial type of perturbation. 
These master equations, of course, can be applied to the perturbation 
analysis of higher-dimensional Schwarzschild 
black holes. Let us mention, furthermore, that these 
master equations 
have been derived in the framework of the gauge-invariant 
formalism for perturbations, and for this reason, they do not 
involve the problem of the choice of gauge.

The purpose of this paper is to show that higher-dimensional 
Schwarzschild black holes are stable with respect to linear perturbations. 
As mentioned above, we have already obtained the master equation 
for each tensorial type of perturbation in the Schwarzschild 
background. Our main focus in this paper is therefore 
to prove that there exist no unstable solutions 
to the master equation with physically acceptable initial data. 
We also show that there exist no well-behaved static 
perturbations that are regular everywhere outside the event horizon.   
This is consistent with the uniqueness\cite{GibbonsIdaShiromizu2002a} 
of a certain class of black holes in an asymptotically flat 
background.
Although our main interest is in the 
standard higher-dimensional Schwarzschild black holes, that is, 
asymptotically flat,\footnote{ 
It has been pointed out, however, that 
the concept of asymptotic flatness itself involves 
some subtle problems 
in higher dimensions, especially in the case of an odd number 
spacetime dimensions\cite{HollandsIshibashi2003a}.  
In this paper, however, we do not discuss this issue  
and do not distinguish the cases of odd and even numbers of 
spacetime dimensions.  
} 
spherically symmetric, static vacuum black holes 
in higher dimensions, we also investigate the stability of 
other maximally symmetric black holes with a non-vanishing cosmological 
constant with respect to tensor- and vector-type perturbations. 
We also give a stability criterion for generalised 
static black holes considered by 
Gibbons and Hartnoll\cite{GibbonsHartnoll2002}.  

In the next section, we first give a general explanation 
of our background metrics, our formalism for gravitational
perturbations, and the basic strategy of our stability analysis. 
At this stage, our arguments are not limited to the case 
of Schwarzschild black holes, 
but are equally applicable to any case of maximally symmetric 
black holes with a non-vanishing cosmological constant. 
Then, we perform the stability analysis for tensor, 
vector and scalar perturbations separately. 
In \S~\ref{sect:static}, we examine static scalar-type 
perturbations in the Schwarzschild background. 
Section~\ref{sect:summary} is devoted to summary and discussion.

\section{Stability analysis} 
\label{sect:Stability Analysis}

As our background, we consider the $(n+2)$-dimensional 
metric of the form   
\bena
 ds^2 = -f(r)dt^2 + \frac{dr^2}{f(r)} + r^2 d\sigma^2_{n} \,,  
\label{metric:Schwarzschild}
\eena  
where 
\bena 
 f(r) = K - \frac{2M}{r^{n-1}} - \lambda r^2 \,, 
\label{fnc:Schwarzschild-lapse}    
\eena 
with constant parameters $M$, $\lambda$ and $K=0,\pm 1$, 
and where $d \sigma_n^2 = \gamma(z)_{ij}dz^idz^j$ denotes the metric 
of the $n$-dimensional maximally symmetric space ${\cal K}^n$ 
with sectional curvature $K=0,\pm 1$.\footnote{
Accordingly, we also refer to the metric~(\ref{metric:Schwarzschild})  
as a {\sl maximally symmetric black hole} 
One can also construct static, vacuum black hole metrics 
by simply replacing $d \sigma_n^2= \gamma_{ij}dz^idz^j$ 
in (\ref{metric:Schwarzschild}) by an arbitrary metric 
of an $n$-dimensional Einstein 
manifold\cite{Birmingham1998,GibbonsHartnoll2002, 
GibbonsHartnollPope2003} whose Ricci curvature 
is $\hat{R}_{ij}=K(n-1)\gamma_{ij}$. 
We briefly comment on such a generalised black hole case 
in \S~\ref{sect:summary}.  
} 
The constant parameter $\lambda$ is related to the $(n+2)$-dimensional 
cosmological constant $\Lambda$ as $\lambda = 2\Lambda/n(n+1)$, 
and the parameter $M$ defines the black hole mass. 
For example, in the case $K=1$ and $\lambda =0$, 
this mass is given by  
\bena
   \frac{ n M {\cal A}_{n} }{8 \pi G_{n+2}},  
\eena 
where ${\cal A}_n = 2\pi^{(n+1)/2}/\Gamma[(n+1)/2]$ is 
the area of a unit $n$-sphere, and $G_{n+2}$ denotes the $n+2$-dimensional 
Newton constant. We assume that 
$M\ge 0$ to avoid the appearance of a naked singularity 
in our background. 
We are interested in the static 
region---called the {\sl Schwarzschild wedge}---in which $f(r)>0$,  
so that $t$ is a time-like coordinate. 
Thus, when $\lambda \ge 0$, the sectional curvature $K$ must be
positive, while when $\lambda <0$,  
$K$ can be $0$ or $\pm 1$. 
The higher-dimensional Schwarzschild black hole 
corresponds to the case $K=1,\,M > 0$ and $\lambda =0$.  

To study perturbations in this background, it is convenient 
to introduce harmonic tensors in the $n$-dimensional symmetric 
space~$({\cal K}^n,d\sigma_n^2)$ and expand the perturbations 
in terms of them, as we see below. 
The type of the harmonic tensor defines the type of the perturbation. 
Combining the expansion coefficients, 
one can construct the gauge-invariant variables in the two-dimensional 
orbit spacetime spanned by the coordinates $y^a = (t,r)$. 
The linearised Einstein equations then reduce to a set of equations 
for the gauge-invariant quantities. 
It is found from the Einstein equations that there is a master 
scalar variable $\Phi$ for each tensorial type of perturbation. 
Consequently, the linearised Einstein equations reduce to a set of 
three master equations for the master variable. 
Thorough studies of the gauge-invariant formalism for perturbations 
and the derivation of the master equations are given in
Refs.~\citen{KI2003} and~\citen{KIS2000}.  

Because our background metric is static, with a Killing parameter $t$ 
in the Schwarzschild wedge, the master equations can be cast into 
the form   
\bena 
 \frac{\partial^2}{\partial t^2} \Phi
   =  \left(\frac{\partial^2}{\partial r_*^2} - V \right) \Phi \,,  
\label{eq:A}
\eena 
where $r_* \equiv \int dr f^{-1}(r)$ and $V$ is a smooth 
function of $r_*$.  
Our task is then to show that the differential operator    
\bena   
 A = - \frac{d^2}{dr_*^2} + V  
\label{ope:A}
\eena
is a positive self-adjoint operator in $L^2(r_*,dr_*)$,  
the Hilbert space of square integrable functions of $r_*$, 
so that Eq.~(\ref{eq:A}) possesses no negative mode solutions 
that are normalisable. %
In other words, the positivity of the self-adjoint operator ensures that, 
given well-behaved initial data---for instance,   
smooth data of compact support in $r_*$---$\Phi$ remains bounded 
for all time. A rigorous proof of the stability of the system 
described by Eq.~(\ref{eq:A}) has been given, in the language of 
spectral theory, by Wald~\cite{Wald1978}. 

It is known that, in the four-dimensional case, 
$V$---called the Regge-Wheeler potential for axial 
perturbations and the Zerilli potential for polar 
perturbations---is positive definite. Hence, the positivity 
of the self-adjoint operator immediately follows. 
However, in systems of greater than four dimensions, 
$V$ is, as we will see, in general not positive definite.      
Thus it is not clear whether $A$ can be extended to a positive 
self-adjoint operator. 
However, because $V$ is bounded from below in most cases, 
in the domain $C^\infty_0(r_*)$---the set of smooth functions 
with compact support in $r_*$---$A$ is a symmetric, semi-bounded 
operator and hence can be extended to a semi-bounded self-adjoint 
operator in such a way that the lower bound of the spectrum of 
the extension is the same as that of $A$. 
In fact, a natural extension, called the Friedrichs extension\cite{RS1975}, 
is such a self-adjoint operator.   
Furthermore, when $\lambda \ge 0$, there is a unique 
self-adjoint extension because in this case, the Schwarzschild wedge 
is globally hyperbolic, and the range of $r_*$ is complete, i.e., 
$(-\infty,\infty)$. Therefore 
the time evolution of $\Phi$---which is defined by a self-adjoint 
extension of $A$---must be uniquely determined by its initial data,  
by virtue of this global hyperbolicity.  
(If there were different self-adjoint extensions, then 
each of them would describe a different time evolution of $\Phi$ 
from the same initial data.) 
Therefore, if the symmetric operator $A$ with domain $C^\infty_0(r_*)$ 
is shown to be positive definite, then the self-adjoint extension must
also be positive definite. We show that this is indeed the case.     

When $\lambda<0$, the situation is quite different,  
because the range of $r_*$ is not complete, being $(-\infty,0)$,   
and the Schwarzschild wedge is no longer globally hyperbolic.  
Nevertheless, there is a prescription\cite{IshibashiWald2003a} 
for defining the dynamics 
of $\Phi$ in terms of a self-adjoint extension of $A$. 
We choose the self-adjoint extension that is obtained by imposing 
the Dirichlet boundary condition $\Phi =0$ at $r_* =0$ 
as a natural requirement for perturbations 
that are well-behaved (in $r_*$). 
The other possible choices of the boundary conditions at $r_* =0$ 
and the corresponding self-adjoint extensions 
in anti-de Sitter spacetime 
will be discussed elsewhere\cite{IshibashiWald2003b}.  

In the subsequent sections, we examine tensor, vector  
and scalar perturbations separately, 
briefly reviewing the metric perturbations and the master equation.

\subsection{Tensor perturbations} 
\label{sect:tensor} 

We begin by considering tensor perturbations, which are given by 
\bena
 \delta g_{ab}=0, \quad 
 \delta g_{ai}=0,\quad  
 \delta g_{ij}=2r^2 H_T \tensor_{ij} \,,  
\label{metricexpansion:tensor}
\eena 
where $H_T$ is a function of $y^a = (t,r)$, and the 
harmonic tensors $\tensor_{ij}$ are defined as solutions to 
the eigenvalue problem on the $n$-sphere, i.e., 
\bena
(\hat\triangle_n + k_T^2)\tensor_{ij}=0 \,,
\eena
with    
\bena
 \tensor^i{}_i=0, \quad \hat D_j \tensor^j{}_i=0 \,.     
\eena
Here, $\hat D_j$ is the covariant derivative with respect to 
the metric $\gamma_{ij}$ 
and $\hat\triangle_n \equiv \gamma^{ij} \hat D_i \hat D_j$.  
The eigenvalues $k_T^2$ are given by $k_T^2 = l(l+n-1)-2$ for $K=1$ 
and form a continuous set of non-negative values for $K \le 0$.  
Note that in the expansion~(\ref{metricexpansion:tensor}), 
we have omitted the indices ${\bf k}$ labelling  
the harmonics and the summation over them. 
Note also that, in the four-dimensional spacetime ($n=2$), 
there are no corresponding modes, as there are no harmonic 
tensors of this type on a $2$-sphere.  

It can immediately be seen that the expansion coefficient $H_T$ is 
itself gauge independent, and it can be taken as the master 
variable for a tensor perturbation.   
The master equation follows from the vacuum Einstein 
equations [Eq.~(5.5) in Ref.~\citen{KI2003} 
or, equivalently, Eq.~(137) in Ref.~\citen{KIS2000}] 
\bena 
 \left( 
       \Box - \frac{V_T}{f} 
 \right) \Phi =0 \,, 
\label{eq:covariant-master} 
\eena  
where $\Phi = r^{n/2}H_T$, and $\Box$ denotes 
the d'Alembertian operator in the two-dimensional orbit 
spacetime with the metric $g_{ab} dy^ady^b = -f(r)dt^2 + f^{-1}(r)dr^2$. 

\begin{figure}
   \centerline{\includegraphics[width=8 cm,height=9 cm]{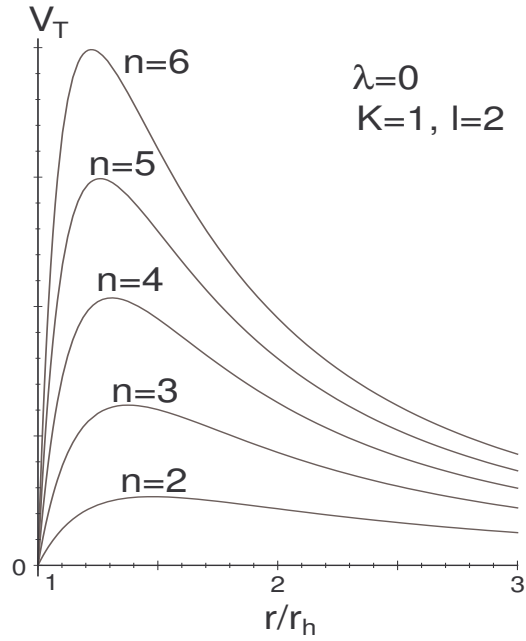}}
   \caption{The potential barriers $V_T$ surrounding 
    $(n+2)$-dimensional Schwarzschild black holes 
    for tensor perturbations of the mode $l=2$.
    Here, $r_h$ denotes the horizon radius.} 
   \label{fig:PotTL0l2}
\end{figure}

\begin{figure}
   \centerline{\includegraphics[width=8 cm,height=9 cm]{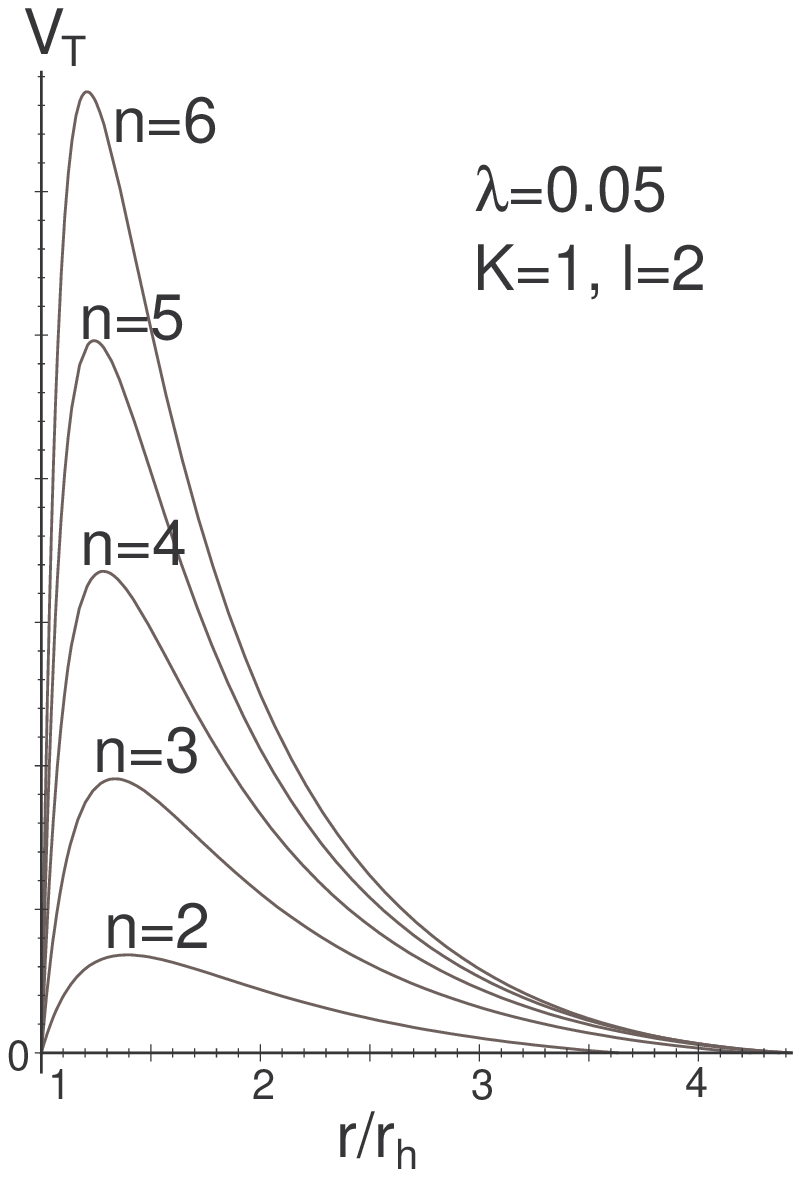}}
   \caption{The potential barriers $V_T$ surrounding 
    $(n+2)$-dimensional Schwarzschild-de Sitter black holes 
    for tensor perturbations of the mode $l=2$. The horizontal axis 
    is normalised with respect to the black hole horizon radius $r_h$.} 
   \label{fig:PotTPLl2}
\end{figure}

\begin{figure}
   \centerline{\includegraphics[width=8 cm,height=9 cm]{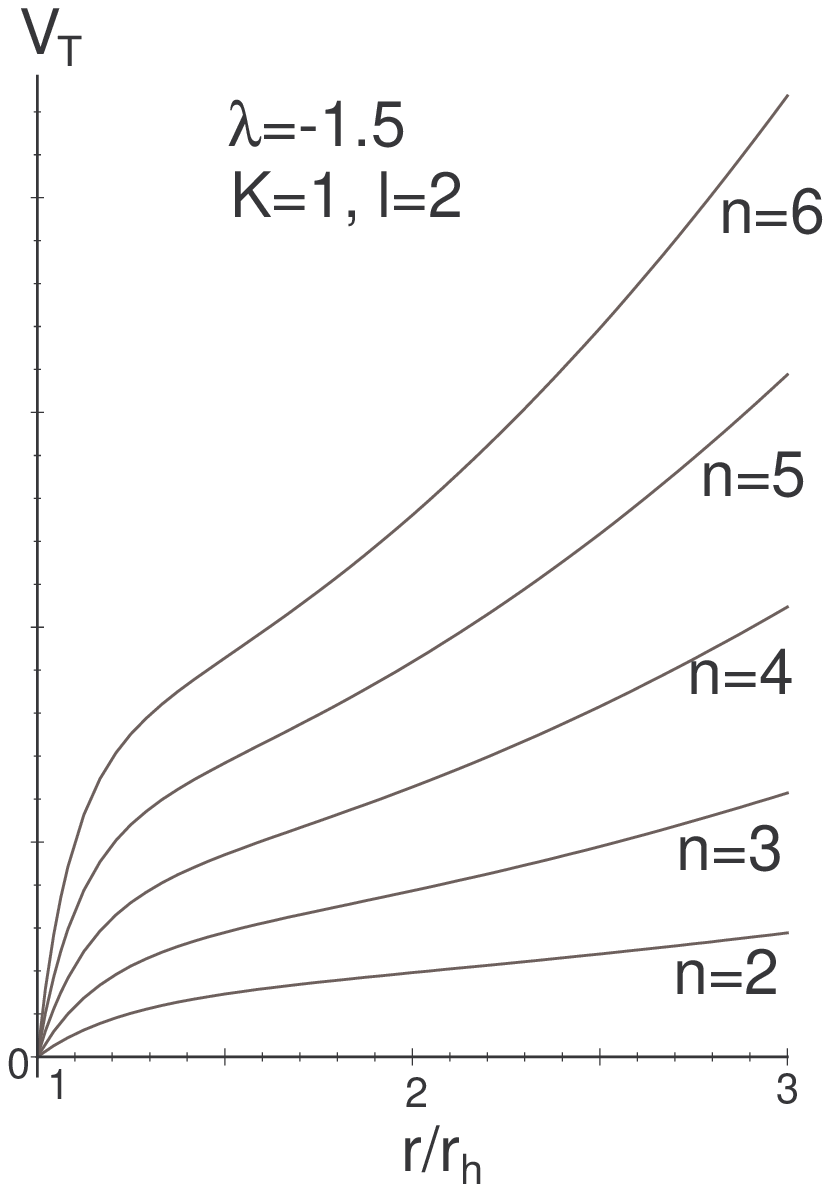}} 
   \caption{The potential barriers $V_T$ surrounding 
    $(n+2)$-dimensional Schwarzschild-anti-de Sitter black holes 
    for tensor perturbations of the mode $l=2$.The horizontal axis 
    is normalised with respect to the black hole horizon radius $r_h$.}
   \label{fig:PotTNLl2}
\end{figure}

The master equation~(\ref{eq:covariant-master}) can be written in 
the form of Eq.~(\ref{eq:A}), with 
\bena 
 V=V_T \equiv \frac{f}{r^2}  
         \left[
                \frac{n(n+2)}{4}f+\frac{n(n+1)M}{r^{n-1}}+k_T^2-(n-2)K 
         \right]  \,. 
\label{Tensorpotential}    
\eena 
For the case of the standard Schwarzschild black hole 
($K=1,\lambda =0$), 
the asymptotic behaviour of $V_T$ at large distances 
($r_* \rightarrow \infty$) 
and near the horizon ($r_* \rightarrow - \infty$) is 
\bena
 V_T \approx \frac{C_+ }{r^2}   \quad \mbox{for} \quad 
      r_* \rightarrow +\infty \,, \quad 
 V_T \approx C_- \exp \left(2\kappa r_*\right) \quad \mbox{for} \quad 
       r_* \rightarrow -\infty \,, 
\label{behaviour:potential} 
\eena  
where $\kappa = (n-1)/2r_h$, with $r_h^{n-1} \equiv 2M$,  
and $C_\pm $ are positive constants. 
Typical profiles of the potential $V_T$ are plotted in 
Figs.~\ref{fig:PotTL0l2}, \ref{fig:PotTPLl2}, 
and \ref{fig:PotTNLl2}.  
We immediately see that the potential $V_T$ is positive definite 
in the Schwarzschild wedge,  
irrespective of the signs of $K$ and $\lambda$ 
[noting that, for $K=1$, we have $k_T^2 - (n-2)K = (l-1)(l+n) >0$]. 
Thus, the operator $A$ with the domain $C^\infty_0(r_*)$ is positive 
and symmetric (Hermitian) with respect to the inner product  
\bena
 (\Phi_1, \Phi_2)_{L^2} 
 &\equiv& \int 
              dr_* \Phi_1^*(r)\: \Phi_2(r) \,.   
\label{def:innerproduct}  
\eena  
Therefore, for the case $r_*\in (-\infty,\infty)$ (that is,  
the case in which $\lambda \ge 0$), $A$ is extended to a unique  
positive, self-adjoint operator in $L^2(r_*)$.  
For the case $\lambda <0$, $r_*$ has the range $(-\infty, 0)$. 
For $V=V_T$, however, $A$ is of the limit-point type\footnote{ 
This type is characterised by the condition that for both $\pm i$, 
one of the solutions to $(-\partial^2/\partial r_*^2 + V)\Phi=\pm i\Phi$ 
is not square integrable at $r_*=0$. 
If $A$ is of the limit-point type at both boundaries, 
then $A=-d^2/dr_*^2 + V$ is essentially self-adjoint. 
This can be seen by noting that 
$V_T \approx n(n+2)/r_*^{2} > 3/(4r_*^{2})$ near $r_*=0$\cite{RS1975}.     
}                
at $r_*=0$, and hence $A$ is uniquely extended to a positive, 
self-adjoint operator. We therefore conclude that 
not only the standard Schwarzschild black hole ($K=1,\lambda =0$) 
but all the possible maximally symmetric black holes 
in higher dimensions are stable with respect to tensor perturbations.  
This result agrees with that of the previous 
analysis\cite{GibbonsHartnoll2002}.

\subsection{Vector perturbations} 
\label{sect:vector} 

Next, consider vector perturbations, which are given by 
\bena
 \delta g_{ab}=0 \,, \quad 
 \delta g_{ai}=r f_a \vector_i ,\quad 
 \delta g _{ij} = -\frac{2}{k_V}r^2 H_T {\hat D}_{(i}\vector_{j)} \,, 
\eena 
where the vector harmonics are introduced as the solutions of 
\bena
 (\hat\triangle_n + k_V^2) \vector_i=0 \,, \quad 
  \hat D_i \vector^i=0 \,, 
\eena 
with eigenvalues $k_V^2$ that are given by 
$k_V^2=l(l + n -1) -1$ with $l\ge 1$ for $K=1$ and 
form a continuous set of non-negative values for $K\le 0$. 

A gauge-invariant variable $F_{a}$ can be constructed by 
combining the expansion coefficients $f_a$ and $H_T$.  
Substituting this into the Einstein equations, 
we can obtain the master scalar variable $\Phi$, 
in terms of which the gauge-invariant variable of 
vector perturbations for $k_V^2>(n-1)K$ is given by 
\bena 
   F_a = f_a + \frac{r}{k_V} D_a H_T 
       = r^{-(n-1)} \epsilon_{ab} D^b (r^{n/2} \Phi) \,,
\eena   
where $\epsilon_{ab}$ denotes the anti-symmetric tensor 
on the two-dimensional orbit spacetime. 
Note that for $K=1$ and $l=1$, $\vector_i$ becomes the Killing 
vector on the $n$-sphere. In this case, only perturbations 
corresponding to rotations of the background geometry satisfy the 
Einstein equations, as explained in Ref.~\citen{KIS2000}. 
Therefore we do not consider the $l=1$ mode for $K=1$ in what follows. 

\begin{figure}
   \centerline{\includegraphics[width=8 cm,height=9 cm]{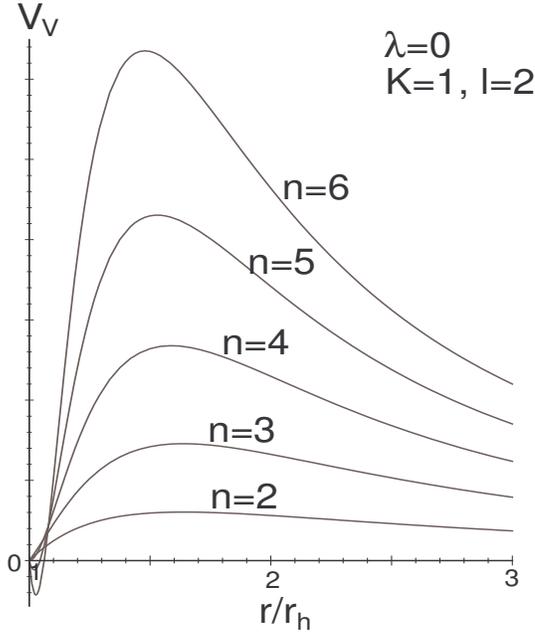}}
   \caption{The potential barriers $V_V$ surrounding 
    $(n+2)$-dimensional Schwarzschild black holes 
    for vector perturbations of the mode $l=2$. The horizontal axis 
    is normalised with respect to the black hole horizon radius $r_h$.
    }
   \label{fig:PotVL0l2}
\end{figure}

\begin{figure}
   \centerline{\includegraphics[width=8 cm,height=9 cm]{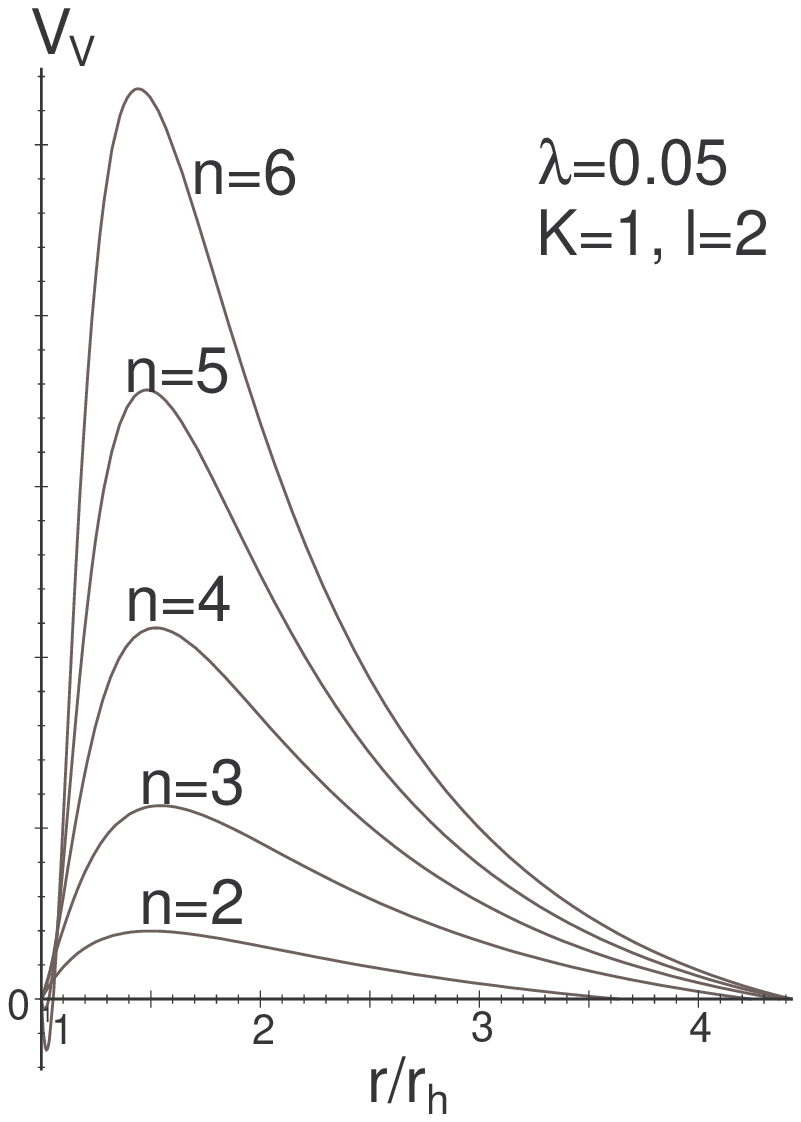}}
   \caption{The potential barriers $V_V$ surrounding 
    $(n+2)$-dimensional Schwarzschild-de Sitter black holes 
    for vector perturbations of the mode $l=2$. The horizontal axis 
    is normalised with respect to the black hole horizon radius $r_h$.}
   \label{fig:PotVPLl2}
\end{figure}

\begin{figure}
   \centerline{\includegraphics[width=8 cm,height=9 cm]{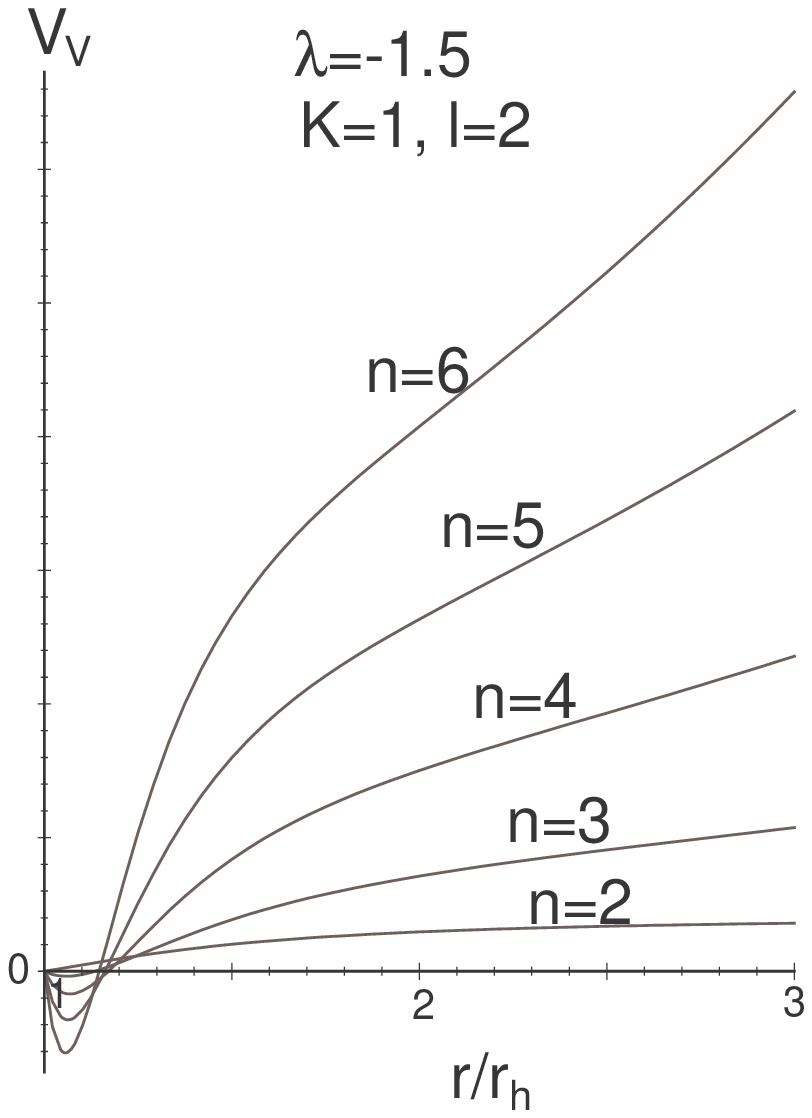}}
   \caption{The potential barriers $V_V$ surrounding 
    $(n+2)$-dimensional Schwarzschild-anti-de Sitter black holes 
    for vector perturbations of the mode $l=2$. The horizontal axis 
    is normalised with respect to the black hole horizon radius $r_h$.} 
   \label{fig:PotVNLl2}
\end{figure}

The evolution part of the Einstein equations can then be reduced to 
[Eq.~(5.14) in Ref.~\citen{KI2003} or, equivalently, Eq.~(141) 
in Ref.~\citen{KIS2000}]  
\bena
    \left( 
       \Box - \frac{V_V}{f} 
 \right) \Phi =0 \,, 
\label{eq:covariant-master-V} 
\eena
which is identical to Eq.~(\ref{eq:A}) with 
\bena 
 V=V_V \equiv \frac{f}{r^2} 
            \left[ 
                  k_V^2 -(n-1)K + \frac{n(n+2)}{4}f  
                 - \frac{n}{2} r \frac{df}{dr} 
            \right] \,.     
\label{RWpotential}
\eena
It can be shown that for the four-dimensional Schwarzschild case 
($n=2$, $K=1$ and $\lambda =0$), 
$V_V$ coincides with the Regge-Wheeler potential. 
Also, it should be noted that the potential $V_V$ falls off at large 
distances and near the horizon in the same manners 
as~(\ref{behaviour:potential}) for $K=1$ and  $\lambda = 0$.  
However, $V_V$ is not positive definite for higher-dimensional  
Schwarzschild black holes with $n>2$, $K=1$ and $\lambda =0$.  
In particular, the constant $C_-$ becomes negative 
when $2n > 2l-1 + \sqrt{12l(l-1)+1}>4$ (for $l \ge2$), as seen in 
Fig.~\ref{fig:PotVL0l2}. 
We can also see that the potential is not positive definite 
for the case of a non-vanishing cosmological constant 
in Figs.~\ref{fig:PotVPLl2} and \ref{fig:PotVNLl2}.


Now, to show the positivity of $A$, 
let us introduce the differential operator   
\bena 
   \tilde D \equiv \frac{d}{dr_*} + S \,, 
\label{Ope:D}   
\eena 
with $S$ being some function of $r$. 
Then, we have 
\bena 
 (\Phi, A\Phi)_{L^2} 
       &=& 
         - \left[
                 \Phi^* \tilde D \Phi 
           \right]_{boundary} 
          + \int dr_* 
                      \left(
                            |\tilde D \Phi |^2 + \tilde V |\Phi|^2
                     \right) \,,
\label{InnerProduct}
\eena   
where 
\bena
  \tilde V \equiv V + f \frac{dS}{dr} - S^2 \,. 
\label{Potential:modified}   
\eena 
{}For $\Phi \in C^\infty_0(r_*)$, the boundary term vanishes. 
Inspecting the form of (\ref{RWpotential}), we find that 
the choice
\bena
  S = \frac{nf}{2r} 
\eena 
yields an effective potential that is manifestly positive: 
\bena
  \tilde V = \frac{f}{r^2}[k_V^2 - (n-1)K] \ge 0 \,.   
\label{effpote:Vector}
\eena 
Here, we note that for $K=1$, $k_V^2 - (n-1)K = (l-1)(l+n)>0$. 
Hence, a symmetric operator $A$ with domain $C^\infty_0(r_*)$ 
is positive definite for any $K$ and $\lambda$ 
(as long as the Schwarzschild wedge exits), 
so is the self-adjoint extension. 
For the case $\lambda <0$, the range of $r_*$ is incomplete. However, 
for the higher dimensional case ($n \ge3$), $V=V_V$ is shown 
to be in the limit-point case at $r_*=0$, and $A$ is uniquely extended 
to a positive, self-adjoint operator. 
For the case of four spacetime dimensions ($n =2$), 
imposing the Dirichlet boundary condition at $r_*=0$, as stated above, 
we can obtain a positive, self-adjoint extension of $A$.   
[Note that under the Dirichlet boundary condition, 
the boundary term in~Eq.~(\ref{InnerProduct}) does not 
contribute to the spectrum.]   
We have therefore confirmed that all the possible maximally symmetric 
black holes are stable with respect to vector perturbations.

\subsection{Scalar perturbations} 
\label{sect:scalar} 

Scalar perturbations are given by 
\bena
 \delta g_{ab}= f_{ab} \scalar, \quad  
 \delta g_{ai}= rf_a \scalar_i  , \quad 
 \delta g_{ij}= 2r^2 (H_L\gamma_{ij}\scalar + H_T \scalar_{ij} ) \,,
\eena
where the scalar harmonics $\scalar$, the associated scalar harmonic 
vector $\scalar_i$, and the tensor $\scalar_{ij}$ are defined by 
\bena
  (\hat \triangle_n + k_S^2 ) \scalar =0 \,, \quad
  \scalar_i = -\frac{1}{k_S} \hat D_i \scalar \,, \quad 
  \scalar_{ij} = \frac{1}{k_S^2} \hat D_i \hat D_j\scalar  
                 + \frac{1}{n}\gamma_{ij}\scalar \,,        
\eena
with the eigenvalues $k_S^2$ given by $k_S^2 = l(l+n-1)$ for $K=1$ 
and forming a continuous set of non-negative values for $K\le 0$. 
By definition, $\scalar_{ij}$ 
is a traceless tensor. 
Using the gauge-dependent vector defined on the orbit spacetime by 
\bena
  X_a = \frac{r}{k_S}\left( f_a + \frac{r}{k_S} D_a H_T \right), 
\eena
we can obtain the gauge-invariant 
variables for the scalar perturbation,  
\bena
   F = H_L + \frac{1}{n} H_T + \frac{1}{r} (D^ar) X_a \,, \quad
   F_{ab} = f_{ab} + D_a X_b + D_b X_a \,. 
\eena
Hereafter, we restrict our attention to modes for which $l\ge 2$,  
because the $l=1$ modes do not correspond to any physical degrees 
of freedom, as discussed in Ref.~\citen{KI2003}.  

We find, after a long calculation, that 
with the following choice of the gauge-invariant master variable
$\Phi$, the linearised Einstein equations can be reduced to 
the form of Eq.~(\ref{eq:A}) [or Eq.~(3.11) in Ref.~\citen{KI2003}]:    
\bena
 && 
    \left(  \Box 
           - \frac{V_S }{f} 
    \right) \Phi = 0 \,, 
\label{eq:covariant-master-S} 
\\
 && \Phi = \frac{n\tilde Z - r(X+Y)}{  r^{n/2 -1} 
                                     \left\{ m
                                            + n(n+1)x /2 
                                     \right \}
                                   } \,,  
\eena 
where here and hereafter 
\bena
 x = \frac{2M}{r^{n-1}} \,,  \quad  m= k_S^2 - nK \,,  
\eena 
and $X$, $Y$ and $\tilde Z$ are defined by 
\bena
  X = r^{n-2}\left(F^t_t -2F \right) \,, \quad 
  Y = r^{n-2}\left(F^r_r -2F \right) \,, \quad 
  \partial_t \tilde Z = - r^{n-2} F^r_t \,.  
\label{defs:XYX}
\eena 
The variables $X$, $Y$ and $\tilde Z$ can be expressed in terms of 
$\Phi$ and its 
derivatives with respect to $D_a$ [see Eq.~(3.12) and Eq.~(3.10) 
in Ref.~\citen{KI2003}]. 

\begin{figure}
   \centerline{\includegraphics[width=8 cm,height=9 cm]{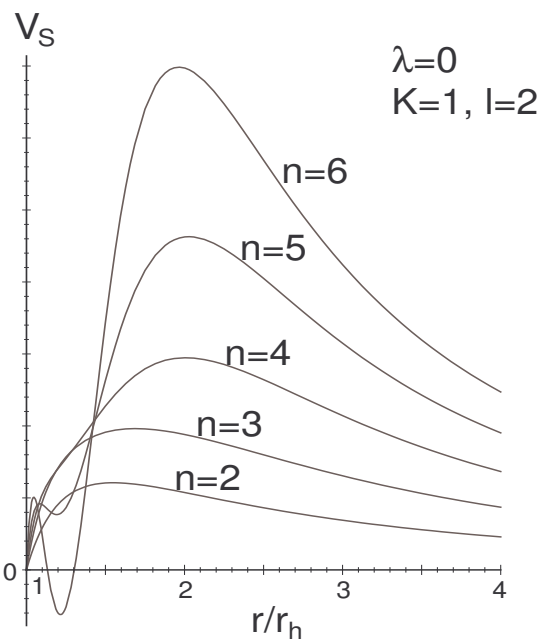}}
   \caption{The potential barriers $V_S$ surrounding 
    $(n+2)$-dimensional Schwarzschild black holes 
    for scalar perturbations of the mode $l=2$. The horizontal axis 
    is normalised with respect to the black hole horizon radius $r_h$.
    In $8$-dimensional spacetimes, $V_S$ is not positive definite.}
   \label{fig:PotSL0l2}
\end{figure}

\begin{figure}
   \centerline{\includegraphics[width=8 cm,height=9 cm]{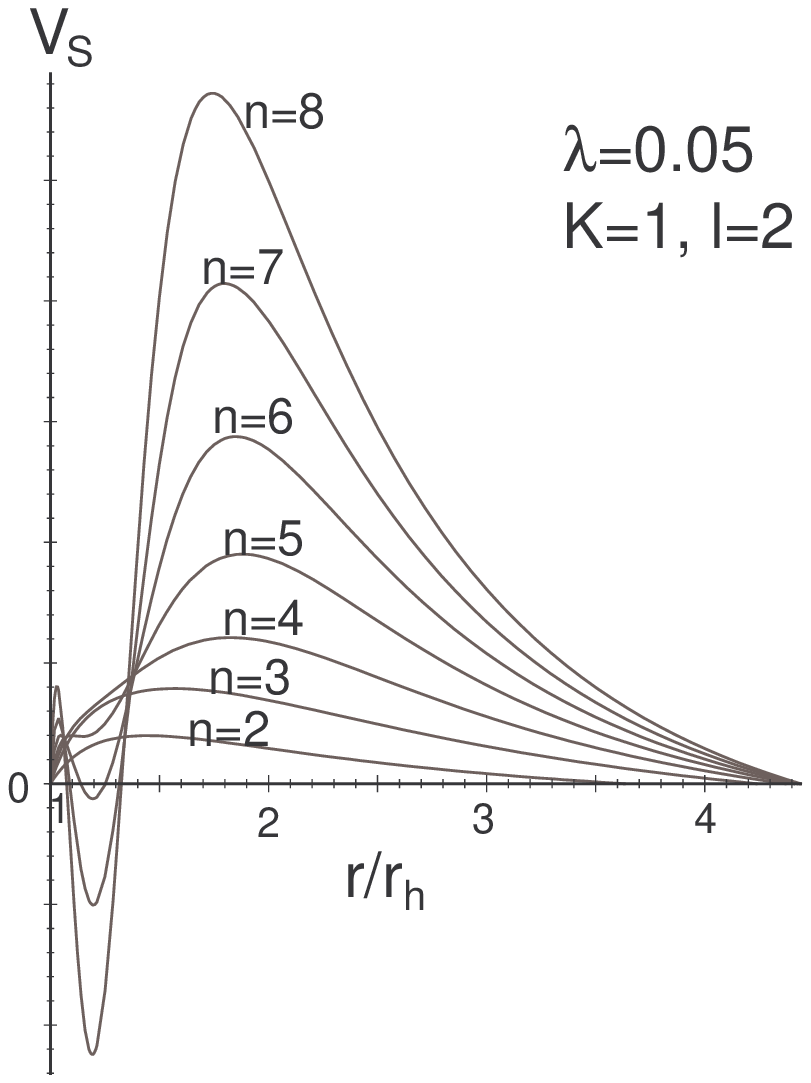}}
   \caption{The potential barriers $V_S$ surrounding 
    $(n+2)$-dimensional Schwarzschild-de Sitter black holes 
    for scalar perturbations of the mode $l=2$. The horizontal axis 
    is normalised with respect to the black hole horizon radius $r_h$.
    In $8$-dimensional spacetimes, $V_S$ is not positive definite.}
   \label{fig:PotSPLl2}
\end{figure}

\begin{figure}
   \centerline{\includegraphics[width=8 cm,height=9 cm]{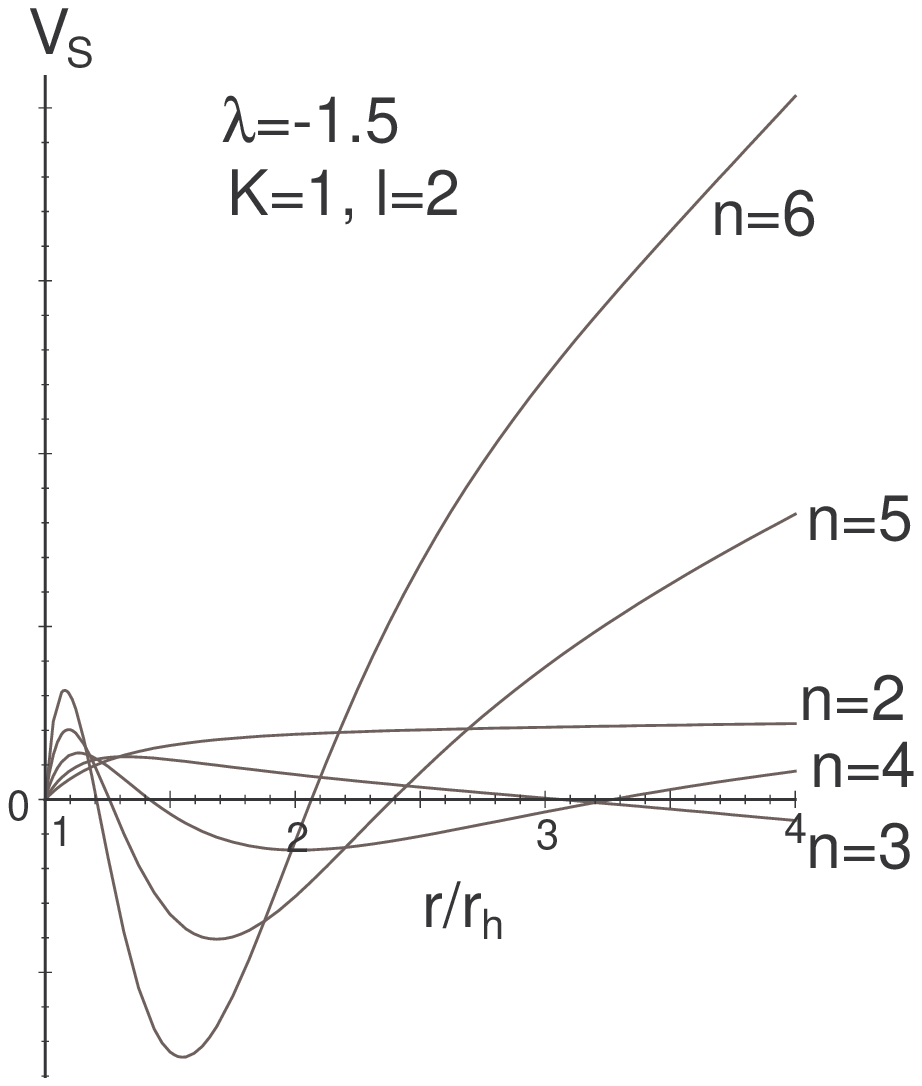}}
   \caption{The potential barriers $V_S$ surrounding 
    $(n+2)$-dimensional Schwarzschild-anti-de Sitter black holes 
    for scalar perturbations of the mode $l=2$. The horizontal axis 
    is normalised with respect to the black hole horizon radius $r_h$. 
    When the number of spacetime dimensions is 
    greater than 4, $V_S$ is not positive definite.}  
   \label{fig:PotSNLl2}
\end{figure}

The potential term in the master equation~(\ref{eq:covariant-master-S}) 
is given by 
\bena
 V_S = \frac{f(r)Q(r)}{16r^2 \left\{ m+n(n+1)x/2 \right\}^2} \,,  
\eena  
where    
\bena 
   Q(r) &=& - \left[
              n^3(n+2)(n+1)^2x^2 -12n^2(n+1)(n-2)mx + 4(n-2)(n-4)m^2 
              \right] \lambda r^2 
\non \\       
        &{}& \, 
            + n^4(n+1)^2x^3  
            + n(n+1)\left[4(2n^2-3n+4)m+n(n-2)(n-4)(n+1)K \right]x^2
\non \\
        &{}& \, 
            -12n\left[(n-4)m+n(n+1)(n-2) \right] mx +16m^3+4n(n+2)m^2 .
\eena 
The manners in which $V_S$ falls off at large distances and near the
horizon are the same as in the vector perturbation case. 
It can be shown that the potential $V_S$ coincides with the Zerilli 
potential and is positive definite in the four-dimensional 
Schwarzschild case ($n=2$, $K=1$ and $\lambda=0$). 
However, for higher dimensions, 
the potential $V_S$ is not positive definite in general,  
as shown in Fig.~\ref{fig:PotSL0l2}. (For further reference, 
we also plot the potential $V_S$ for the Schwarzschild-de Sitter 
case in Fig.~\ref{fig:PotSPLl2} and 
the Schwarzschild-anti-de Sitter case in Fig.~\ref{fig:PotSNLl2}.) 

{}From this point, we focus on the standard higher-dimensional Schwarzschild 
black hole case, i.e., that with $\lambda =0$ and $K=1$. 
Taking $S$ to be 
\bena
  S = \frac{f}{h}\frac{dh}{dr} \,, \quad 
  h \equiv r^{n/2+l-1} \left\{ 
                              (l-1)(l+n) + n(n+1)x /2  
                       \right\} \,,     
\eena 
we obtain as the effective potential defined 
in Eq.~(\ref{Potential:modified}) the form      
\bena
   \tilde V = \frac{f(r) \tilde Q(r) 
                    }{
                      2r^{2} \left\{(l-1)(l+n) + n(n+1) x/2 \right\}  
                     }\,, 
\eena  
where   
\bena
 \tilde Q(r) \equiv lx[ ln (n+1)x +2(l-1)\{n^2+n(3l-2) + (l-1)^2\}] \,.
\eena
Clearly $\tilde V > 0$, and thus $A$ with domain $C^\infty_0(r_*)$  
is strictly positive, so is the self-adjoint extension. 
The higher-dimensional Schwarzschild black holes 
are therefore stable with respect to scalar perturbations. 
This completes the stability analysis for higher-dimensional 
Schwarzschild black holes.

\section{Static perturbations} 
\label{sect:static} 

In this section, we treat ``static'' scalar perturbations  
in the background of the Schwarzschild metric, 
with $K=1$, $\lambda =0$ and $n\ge 2$. 
Because scalar perturbations represent deviations 
from a spherical form, if there exist any regular static 
perturbations that fall off at large distances in such a manner 
that asymptotic flatness is maintained, it would be implied that 
there exist asymptotically flat, static vacuum black holes 
whose event horizons are topologically $S^n$ 
but are not spherical. We show that this is not the case.


Regarding $Y$ in Eq.~(\ref{defs:XYX}) as the master variable 
for a static perturbation, we have the following 
master equation for $Y$ 
[with the replacement of $r$ by $x=2M/r^{n-1} \equiv (r_h/r)^{n-1}$ 
in Eq.~(4.4) of Ref.~\citen{KI2003}]:  
\bena 
  \left[ 
         \f(d^2/dx^2) 
        + \frac{2(n-2)x+2}{(n-1)x(x-1)} \f(d/dx) 
        + \frac{-(l-1)(l+n)+(n-2)x}{(n-1)^2x^2(1-x)} 
  \right] Y = 0 \,.   
\label{eq:Y}
\eena  
There are general solutions to Eq.~(\ref{eq:Y}) that can be expressed 
in terms of hyper-geometric functions as 
\bena
&&
  Y = C_1 \Frac(r_h/r)^{-(l-1)}  
         F \left(   
                \f(n-1-l/n-1), -\f(l/n-1), -\f(2l/n-1); x         
           \right)
\non\\ 
&&{} \qquad 
    + C_2 \Frac(r_h/r)^{l+n}
         F \left(   
                \f(n-1+l/n-1), -\frac{2(n-1)+l}{n-1}, 
               -\frac{2(n+l-1)}{n-1}; x           
           \right) \,. 
\eena 
{}From this expression, we find that the original gauge-invariant 
variables $F$ and $F^a_b$ behave at the spatial infinity as $pC_1 
r^l+q C_2/r^{n+l-1}$, where $p$ and $q$ are non-vanishing 
constants that depend on the variables. 
Hence, in studying the question of whether regular, 
well-behaved static perturbations exist or not, 
we can choose the boundary condition $C_1 =0$, 
so that the gauge-invariant variable 
$Y$ for $l\ge 1$ vanishes as $r_*\rightarrow \infty$ in the form 
\bena
  Y \sim O\Frac(1/r^{l+n}) \,. 
\eena 
Note that the $l=0$ modes preserve the spherical symmetry of 
the black hole, and hence need not be considered here. 

Now, we redefine the master variable as  
\bena
  \tilde Y \equiv \sqrt{ \frac{f}{r^n} } \frac{r}{f'} Y \,,       
\eena 
for which the asymptotic condition is written  
\bena 
  \tilde Y \sim O \Frac(1/r^{n/2+l-1}) \,. 
\label{behaviour:tildeY}
\eena   
Then, we can show that $\tilde Y$ obeys [see Eq.~(4.8) 
in Ref.~\citen{KI2003}] the equation 
\bena
  A \tilde Y = \left( 
                     - \frac{d^2}{dr_*^2} + \tilde V 
               \right) \tilde Y = 0 \,,    
\label{eq:zeromode}
\eena
with the potential [see Eq.~(4.11) in Ref.~\citen{KI2003}]:  
\bena
 \tilde V &=& \frac{n(n-2) + 4 l(l+n-1)(1-x) + 2nx(1 -x) + x^2}{4r^2} \,.   
\eena 
Because this potential is positive definite, we can show that there
exists no solution that satisfies the asymptotic
condition~(\ref{eq:zeromode}) and
is bounded at the horizon. First, from Eq.~(\ref{eq:zeromode}), 
there is a point $r_*=a$
at which $\tilde Y$ is positive and $d\tilde Y/dr_*=C<0$, 
by changing the sign of $\tilde Y$ if necessary. 
Suppose that $\tilde Y$ is not positive definite in $r<a$. 
Then, there must be a point $r_*=b<a$ such that $\tilde Y$ 
is positive in the interval $b<r_*<a$, and vanishes 
at $r_*=b$. However, this cannot happen because $d^2\tilde Y/dr_*^2>0$ 
in $b<r_*<a$ and $d\tilde Y/dr_*<0$ at $r_*=a$. Hence,
$\tilde Y$ is always positive in $r_*<a$. From this and the positivity of
$\tilde V$, it follows that $d\tilde Y/dr_*\le C$ in $r_*<a$. Hence,
$\tilde Y$ increases without bound as $r_* \rightarrow-\infty$. 
We therefore conclude that there exists no static perturbation 
that falls off at large distances and is regular everywhere outside 
the event horizon.

\section{Summary and discussion} 
\label{sect:summary} 

We have shown that higher-dimensional Schwarzschild black holes 
are stable with respect to all tensorial types of linear perturbations, 
examining the master equations recently derived in the framework 
of the gauge-invariant formalism for black hole 
perturbations\cite{KIS2000,KI2003}. 
We have confirmed that the 2nd-order elliptic differential operator $A$ 
appearing in the master equation is a positive, self-adjoint operator 
in the Hilbert space $L^2$, and we have thereby shown that each type of 
master equation possesses no negative-mode solutions 
that are normalisable. 
We have also proved that there exists no regular, well-behaved 
static scalar perturbation of Schwarzschild black holes. 
This result is consistent with the uniqueness of higher-dimensional 
spherically symmetric, static vacuum black holes.  

Our strategy explained in \S~\ref{sect:Stability Analysis} 
is not limited to the standard Schwarzschild black hole case.  
In fact, with the same method, we were also able to show 
the stability of maximally symmetric black holes 
for any $K$ and $\lambda$ 
(and thus, including the higher-dimensional Schwarzschild-de Sitter 
and Schwarzschild-anti-de Sitter black holes) with respect to 
tensor and vector perturbations. 
Thus, to complete the stability analysis of maximally symmetric 
black holes, it is now necessary only to consider the case of 
scalar perturbations of non-asymptotically flat black holes. 
(Note that in the four-dimensional case, the stability 
of Schwarzschild-(anti-)de Sitter black holes has already been 
proved\cite{KI2003}.) 


The background metric~(\ref{metric:Schwarzschild}) describes 
a generalised black hole when $d\sigma_n^2$ is the metric of 
a generic Einstein manifold 
with Ricci curvature $\hat R_{ij} = K(n-1)\gamma_{ij}$. 
The stability of such a ``black-Einstein'' has recently been 
studied by Gibbons~{\em et al.}\cite{GibbonsHartnoll2002, %
GibbonsHartnollPope2003} and Hartnoll\cite{Hartnoll2003}. 
As a reason to concentrate on the analysis of tensor perturbations, 
they argued that a generalised black hole should be 
stable with respect to vector and scalar perturbations because 
the standard higher-dimensional Schwarzschild black hole is believed 
to be stable with respect to these perturbations, at least 
in the case in which the event horizon takes the form of a compact 
Einstein manifold of 
positive curvature. Their argument is based on the following two 
points. First, because the Weyl tensor does not appear in the 
perturbation equations for the scalar and vector perturbations, 
the stability with respect to these modes is determined only by the spectrum 
of the Laplacian on the Einstein manifold. Second, the smallest 
eigenvalues of the Laplacian for these modes are always greater than 
or equal to those for the sphere, and hence the standard Schwarzschild 
black hole is the most unstable. Here, we regard the second smallest 
eigenvalue as the smallest one for the scalar mode, because the $k=0$ 
mode does not affect the stability analysis. Although they proved 
the second claim only for a vector mode, we can also prove it 
for a scalar mode with the help of the identity
\begin{equation}
\int d\sigma_n \SHB^{ij}\SHB_{ij}=\frac{n-1}{nk^2}(k^2-n)
  \int d\sigma_n \SHB_i\SHB^i \ge0.
\end{equation}
Furthermore, the first point implies that 
even for the generalised black hole case, the master equations 
we have examined here are still valid, if we replace the eigenvalues 
$k_V^2$ and $k_S^2$ of the Laplacian on ${\cal K}^n$ by the 
corresponding ones on the Einstein manifold.\footnote{ 
The authors thank Sean Hartnoll for discussion of this 
point\cite{Sean}.} 
Our results in this paper thus place their analysis on a firmer basis, 
giving a proof for their argument summarised above for the case $K=1$. 
 
We now discuss a few points regarding the stability of 
a generalised black hole with respect to tensor perturbations. 
In this case, the perturbation 
couples to the Weyl curvature. However, the master equation is still 
valid if we replace $k_T^2$ by $\lambda_L -2nK$, with $\lambda_L$ 
being the eigenvalue of the Lichnerowicz operator on the Einstein 
manifold, as shown by Gibbons and Hartnoll\cite{GibbonsHartnoll2002}. 
Hence, from the positivity of $V_T$ expressed in the form 
\eqref{Tensorpotential}, we obtain a sufficient condition 
for the stability,   
\begin{equation}
\lambda_L\ge(3n-2)K.
\label{condi:first}
\end{equation}
By rewriting $V_T$ as
\begin{equation}
V_T=\frac{f}{r^2}\left[ \lambda_L+\frac{n^2-10+8}{4}K
     -\frac{n(n+2)}{4}\lambda r^2+\frac{n^2 M}{2r^{n-1}} \right],
\end{equation}
we obtain another sufficient condition, 
\begin{equation}
\lambda_L \ge -\frac{n^2-10n+8}{4}K \,,  
\label{condi:second}
\end{equation}
for $\lambda\le0$. This condition coincides with that derived by 
Gibbons and Hartnoll. Furthermore, if we apply the same trick used for 
vector perturbations to tensor perturbation by setting 
\bena
 S = -\frac{nf}{2r} \,, 
\eena 
we have the inner product~(\ref{InnerProduct}) with the 
effective potential  
\bena 
  \tilde V = \frac{f}{r^2}\left[\lambda_L -2(n-1)K\right] \,. 
\eena
{}From this expression, we obtain the third sufficient condition, 
\begin{equation}
\lambda_L \ge 2(n-1)K.
\label{condi:third}
\end{equation}
For $K=1$ and $\lambda\le0$, the second
condition~(\ref{condi:second}),  
is the strongest, while for $K=1$ and $\lambda>0$,      
the third condition,~(\ref{condi:third}), is the strongest. 
For $K=-1$, the first condition,~(\ref{condi:first}), is the strongest. 

Finally, we note that in the stability analysis carried out 
in the present paper, the supports of 
the initial data for the perturbations are, as one condition 
for being well-behaved, 
restricted to compact regions of the initial hypersurface inside 
the exterior Schwarzschild wedges: We have considered such initial 
data that vanish in an open neighbourhood 
of the event horizon, in particular, on the bifurcation $n$-sphere.  
In the four-dimensional case, this restriction on the initial data 
of the perturbations was removed by Kay and Wald\cite{KayWald1987}. 
The key ingredients of Kay and Wald's theorem are the discrete 
isometry of extended Kruskal spacetime---which is implemented 
even in the higher-dimensional case (and also for cosmological 
horizons)---and the well-posed nature of the Cauchy problem 
for the wave equation---which also holds for the master equations. 
Furthermore, the forms in which the potentials $V$ fall off 
near the horizon and in the asymptotic region are 
the same as in the four-dimensional case. 
Hence, there seems to be no dimensional obstruction 
in applying Kay and Wald's theorem to our present situation,  
which would imply the stronger claim that 
{\sl higher-dimensional Schwarzschild black holes are stable 
with respect to the linear perturbations whose initial data 
have compact support on a ``Cauchy surface'' 
in the extended Kruskal spacetime.}  

\section*{Acknowledgements} 
The authors wish to thank Stefan Hollands and Bob Wald  
for many useful discussions and suggestions at the early stage 
of this work. We also wish to thank Gary Gibbons and Sean Hartnoll 
for discussions and comments, especially on the extension of 
the present analysis to the generalised black hole case. 
A.I. thanks the people of the Enrico Fermi Institute 
for their hospitality. 
This work was supported in part by the Japan Society for the Promotion 
of Science (AI), and by the JSPS grant no.15540267 (HK).

\end{document}